\documentclass[prb,superscriptaddress,amsmath,amssymb,twocolumn]{revtex4-2}
\pdfoutput=1
\usepackage{physics}
\usepackage{lipsum}
\usepackage{soul}
\usepackage{booktabs}
\usepackage{natbib}
\usepackage{float}
 \usepackage{subfloat}
\usepackage[english]{babel}
\makeatother
\usepackage{layouts} 
\usepackage{graphicx}
\usepackage{caption}
\usepackage{subcaption}
\usepackage[usenames,dvipsnames]{color}
\usepackage{stackengine}
\usepackage[T1]{fontenc}
\usepackage[colorlinks=true,allcolors=blue]{hyperref}

\begin{document}

\title{Topological properties of gapless phases in an interacting spinful wire}

\author{Polina Matveeva}
\affiliation{Department of Physics,  University of Minnesota,  Minneapolis,  MN 55455,  USA}
\author{Dmitri Gutman}
\affiliation{Department of Physics, Bar-Ilan University, Ramat Gan, 52900, Israel}
\author{Sam T.~Carr}
\affiliation{School of Engineering,  Mathematics and Physics,  University of Kent, Canterbury CT2 7NH, United Kingdom}

\begin{abstract}
We study topology in gapless phases of an interacting spinful model with spin-charge separation.   We focus on the gapless boundaries between  $\mathbb{Z}_2$ symmetry-breaking phases.  We find two topologically non-trivial gapless states that occur at the boundary between a non-trivial and trivial insulators.  They correspond to topological Luther-Emery liquid and topological Mott insulator.  The Luther-Emery liquid is characterized by gapless charge excitations and features topological edge modes that carry fractional spin,  while the topological Mott insulator has gapless spin sector and features edge states that carry fractional charge.  
Surprisingly even though there is no mean-field description of the interacting gapless phases,  as there is no local order parameter,  we show that they can be adiabatically connected to a non-interacting topological metal.  This non-interacting state is a phase boundary between decoupled Su-Schrieffer-Heeger chains with the winding number $\nu=2$ and chains with $\nu=1$.  
\end{abstract}

\maketitle
\section{Introduction} 

Topological phases of matter are states that feature physical properties that are robust against perturbations.  One of the examples of such properties are the  edge modes of topological insulators.   While bulk of such systems is fully gapped,  there are gapless edge states at the boundary that are protected if certain symmetries are preserved.   The full mathematical classification of the non-interacting phases is completely established and is given in the context of symmetry-protected topological states \cite{Zirnbauer1996,Altland1997,Kitaev2009},  while classification of interacting states is only known for fully gapped one-dimensional models \cite{Turner2011,Wen2009}.   While topological properties are typically discussed in the context of fully gapped states,  one may ask if protected edge modes exist in gapless systems.  A simple non-interacting example of such topological gapless state is a critical phase between the gapped phases of two uncoupled Su-Schrieffer-Heeger (SSH) chains with different topological indices.  At this critical point one of the chains becomes gapless,  while the other remains gapped and topological.  As we will discuss in this paper,  the edge modes in this critical phase are robust with respect to weak perturbations that  preserve symmetry that protects topology of the gapped phases.  More general examples of single-particle topological metals built from Kitaev chains were studied in \cite{Verresen2018},  it was shown that they remain stable against disorder.  

There are many interacting examples of topological metals and superconductors that were studied in different one-dimensional microscopic models.  They include superconductors \cite{Keselman2015,Andrei2020},  one-dimensional wires with spin-orbit coupling \cite{Kainaris2015},  wires with spin-anisotropic interactions \cite{Kainaris2017,Verresen2021},  two interacting chiral modes at the edges of QSH planes \cite{Kainaris2017,Santos2016} and in an effective low-energy model with broken SU(2) symmetry \cite{Kainaris2018}.  All these models have gapless charge excitations and gap in spin degrees of freedom.  
The ground state of such states features topological edge modes that carry fractional spin.  A non-trivial gapless state was also studied in a spin $1/2$ ladder \cite{Jiang2018} where it emerges upon doping a system in the Haldane phase.  There is also an example of a topological Mott insulator studied in \cite{Yoshida2014}.  

In this work,  we investigate in full detail topological gapless states of one-dimensional interacting systems with spin-charge separation,  which is common in such systems.  By studying the full phase diagram of such a model,  we identify  topological gapless phases as transition lines between two different symmetry broken phases.   The gapped sector can be classified analogously to the fully gapped phases discussed in our previous work \cite{Matveeva2025}.  In particular,  we identify phases that feature topologically non-trivial edge states by computing the degeneracy of the edge solitons.  

We then further will show by adding a single particle gap competing with the dynamically generated one that we can adiabatically connect these strongly correlated phases to non-interacting gapless topological phases.  It remains an open question whether this is always possible or not. 

The structure of the paper is as follows.  
In Section  \ref{sec:nonint} we briefly discuss the non-interacting topological metal and its stability.   
In Section \ref{sec:gapless_int} we focus on the gapless states in the interacting model and find two topologically non-trivial phases: topological Luther-Emery liquid (or a topological metal) and a topological Mott insulator. 
Similar phases were identified in Ref. \cite{Montorsi2017} that focused on the same effective bosonic model.  We discuss physical properties of these phases and show that they represent the critical phases between a trivial and topological gapped states with spontaneous symmetry breaking. 
In Section \ref{sec:connection} we show that one can adiabatically deform them to a non-interacting topological metal discussed in Section \ref{sec:nonint}.  
\section{Non-interacting topological metal}
\label{sec:nonint}
\begin{figure}
\includegraphics[scale=0.15]{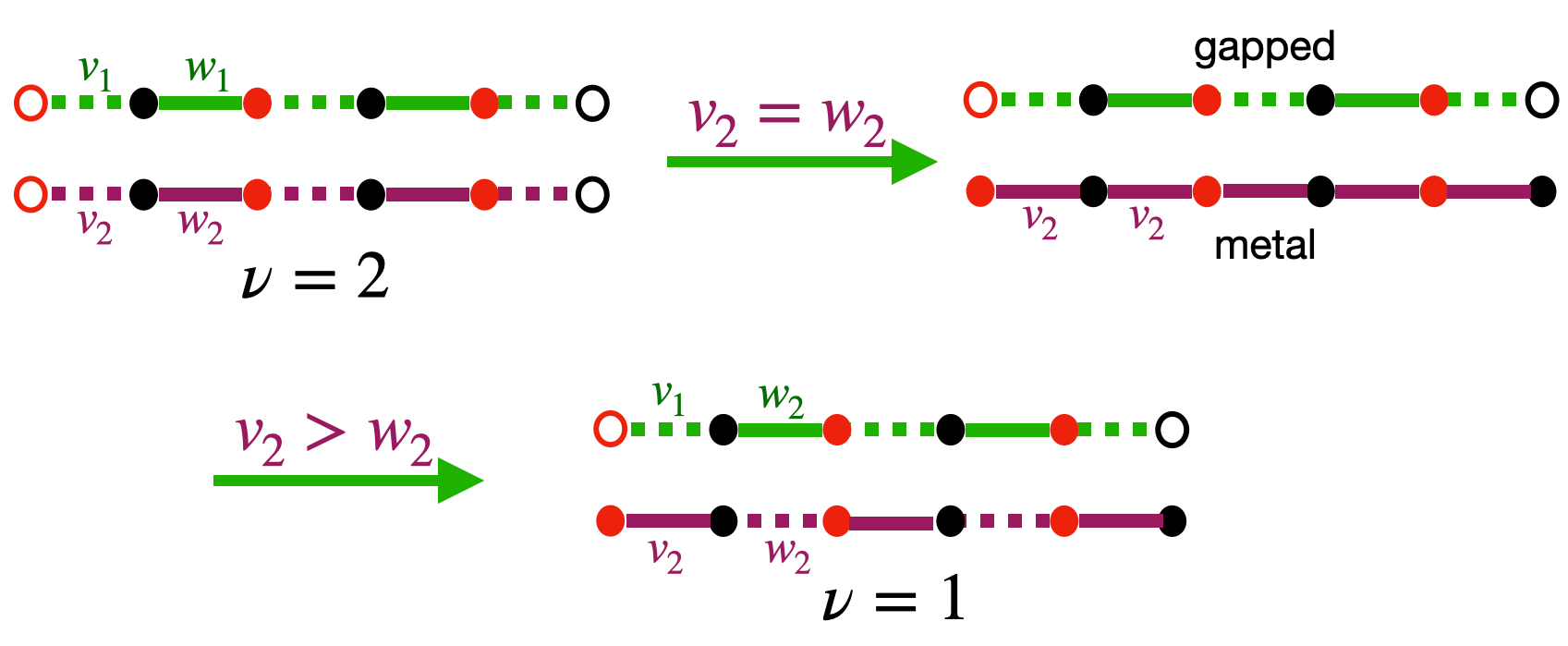}
\caption{Transition between topological phases with different winding numbers in the Su-Schrieffer-Heeger chains.  The critical phase with $w_2=v_2$ represents a topological metal,  the phase where one of the chains is gapless while the second one is gapped and has a zero mode at the boundary.  }
\label{nonint_top_metal}
\end{figure}

Before we discuss topology in the interacting gapless models,  let us consider an example of a non-interacting topological metal.  It occurs at the phase boundary between insulating states with winding numbers $\nu=2$ and $\nu=1$.  The simplest microscopic realization of these insulating phases is given by two decoupled SSH chains described by the following Hamiltonian: 
\begin{align}
\label{SSH_chains}
& H=H_1+H_2,   \nonumber \\
& H_{i} = v_i \sum_{n=1}^N c^{\dagger}_{A,i,n} c_{B,i,n} + w_i \sum_{n=1}^{N-1} c^{\dagger}_{B, i, n} c_{A, i,n+1} +\text{h.c.},
\end{align}
where $i$ denotes a chain index,   $A$ and $B$ are sublattices and $n$ is the index of a unit cell.   For simplicity we assume that all parameters are positive.  In the case $v_i<w_i$ both chains are topologically non-trivial,  so the total winding number is $\nu=2$,  while when only one of the chains is non-trivial,  the total winding number is $\nu=1$.  Now let us tune the parameters of the chains and go from one phase to the other.  Let us start with the case $v_i<w_i$ for both chains and then tune $v_2$ such that the gap in the second chain closes,  that happens at $v_2=w_2$.  At this point the second chain becomes a regular metal,  while the first one remains gapped and topologically non-trivial.   We can then reopen the gap in the second chain such that $v_2>w_2$.  In that case the second chain is topologically trivial,  so the full model has a total winding number $\nu=1$.  This transition is illustrated in Fig.  \ref{nonint_top_metal}.  
The intermediate gapless phase represents a simple non-interacting example of a topological metal,  that features gapless degrees of freedom,  but also has a gapped sector with zero-energy edge modes.  
As we show in Appendix \ref{app:stability},  the zero energy edge modes are stable with respect to perturbations that couple the two chains and preserve the sublattice symmetry.  

\section{Gapless phases in an interacting spinful model}
\label{sec:gapless_int}
\subsection{The model}
\begin{figure}
\includegraphics[scale=0.15]{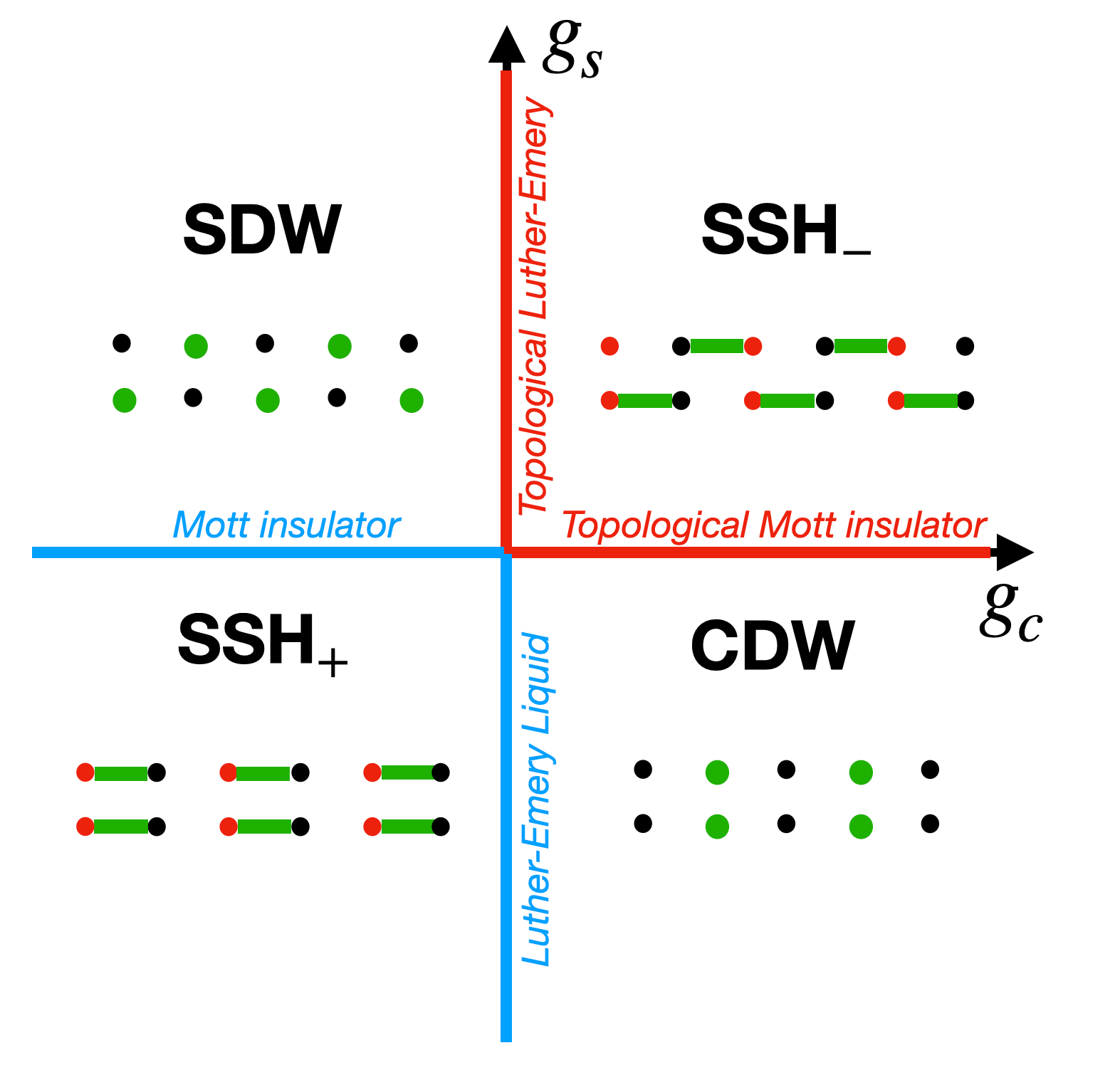}
\caption{Phase diagram of the model \eqref{full_model_main}.  Each of the phases is characterized by an order parameter \eqref{order_parameters_ferm}.  The green dots and lines schematically illustrate the distribution of electronic density.  The topologically non-trivial gapless phases are shown in red,  the trivial ones are shown in blue.  }
\label{phase_diagram}
\end{figure}

Now let us focus on a interacting problem.  We consider the following general bosonic model that describes one-dimensional spinful electrons \cite{gogolin2004bosonization,  giamarchi2003quantum}:  
\begin{align}
\label{full_model_main}
H=H_{\text{LL}} + V. 
\end{align}
Here $H_{\text{LL}}$ is the quadratic part of the Hamiltonian: 
\begin{align}
H_{\text{LL}} =\sum\limits_{\nu =c,s}  \frac{v_{\nu}}{2} \left[K_\nu \Pi^2_{\nu} +K^{-1}_{\nu} (\partial_x \phi_\nu)^2\right], 
\end{align}
where the fields $\phi_c$ and $\phi_s$ describe charge and spin degrees of freedom.  The gap opening $V$ term is given by: 
\begin{align}
\label{gap_terms_main}
V=\frac{g_c}{(\pi a)^2} \cos[\sqrt{8\pi} \phi_c]+\frac{g_s}{(\pi a)^2} \cos[\sqrt{8\pi} \phi_s],
\end{align}
where $a$ is the lattice constant.  
There are many microscopic realizations that lead to this effective model \cite{gogolin2004bosonization,  giamarchi2003quantum},  however in order to reach all possible phases one needs to add interactions that break SU(2) symmetry.   In \cite{Matveeva2025} we discussed such microscopic realization,  and we showed that depending on the signs of the coupling constants $g_c$ and $g_s$ there are four gapped phases that spontaneously break $\mathbb{Z}_2$ symmetry.   
The corresponding order parameters are: 
\begin{align}
\label{order_parameters_ferm}
\mathcal{O}_{\text{CDW}}  &= (-1)^j \sum_{\sigma} c^{\dagger}_{j,\sigma} c_{j,\sigma} \propto
 \nonumber \\ &\propto \sin\left(\sqrt{2\pi} \phi_c\right) \cos\left(\sqrt{2\pi} \phi_s\right),  \nonumber \\
\mathcal{O}_{\text{SSH}_+} &= (-1)^j \sum_{\sigma} \left( c^{\dagger}_{j,\sigma} c_{j+1,\sigma} + \text{h.c.} \right) \propto \nonumber \\
 & \propto \cos\left(\sqrt{2\pi} \phi_c\right) \cos\left(\sqrt{2\pi} \phi_s\right),  \nonumber \\
\mathcal{O}_{\text{SSH}_-} &= (-1)^j \sum_{\sigma,\sigma'} \left( c^{\dagger}_{j,\sigma'} (\sigma_z)_{\sigma'\sigma} c_{j+1,\sigma} + \text{h.c.} \right)
  \propto \nonumber \\ 
  &\propto \sin\left(\sqrt{2\pi} \phi_c\right) \sin\left(\sqrt{2\pi} \phi_s\right),   \nonumber\\
\mathcal{O}_{\text{SDW}}  &= (-1)^j \sum_{\sigma,\sigma'} c^{\dagger}_{j,\sigma'} (\sigma_z)_{\sigma'\sigma} c_{j,\sigma}
  \propto \nonumber \\ &\propto \cos\left(\sqrt{2\pi} \phi_c\right) \sin\left(\sqrt{2\pi} \phi_s\right). 
\end{align}
The order parameters CDW and SDW correspond to charge density and spin density wave phases,  and the parameters $\text{SSH}_+$ and $\text{SSH}_-$  describe phases with bond dimerization.  The phase $\text{SSH}_+$ has the same dimerization patter as two identical SSH chains,  and $\text{SSH}_-$ phase can be deformed to two SSH chains in the opposite topological phases.  These phases are illustrated in Fig.  \ref{phase_diagram}.  We studied topology of these states by computing the degeneracy of the edge modes in a finite model.  We found that $\text{SSH}_-$ is the only phase that features topological edge modes.  

We now turn to the gapless phase boundaries between different gapped phases.  In our model (\ref{gap_terms_main})  they correspond to the lines $g_c=0$ or $g_s=0$,  see the phase diagram in Fig. \ref{phase_diagram}.  Along these lines,  the system is gapless in charge or spin sectors.  The analysis of the edge modes in the gapless phases follows the same approach as in the fully gapped ones.  In particular,  we look at the edge solitons that interpolate between the bulk values of bosonic field in a gapped sector and their boundary values.   In a topologically non-trivial phase the edge modes lead to extra ground state degeneracy in a finite system.  Let us now focus on properties of each gapless state in a detail.

\subsection{Metallic phases - Luther-Emery Liquids}
\begin{figure}
\begin{subfigure}[b]{0.48\columnwidth}
\includegraphics[scale=0.12]{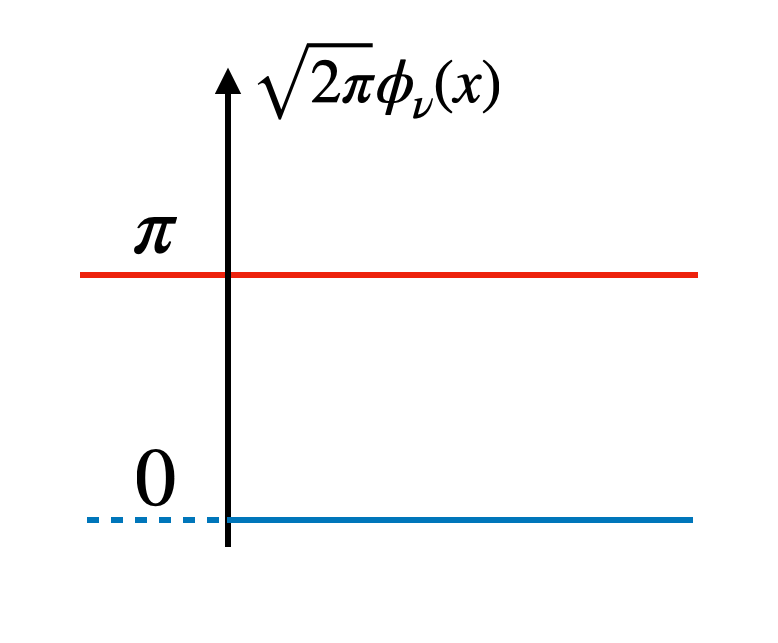}
\caption{}
\label{metal_pic_triv}
\end{subfigure}
\begin{subfigure}[b]{0.48\columnwidth}
\includegraphics[scale=0.11]{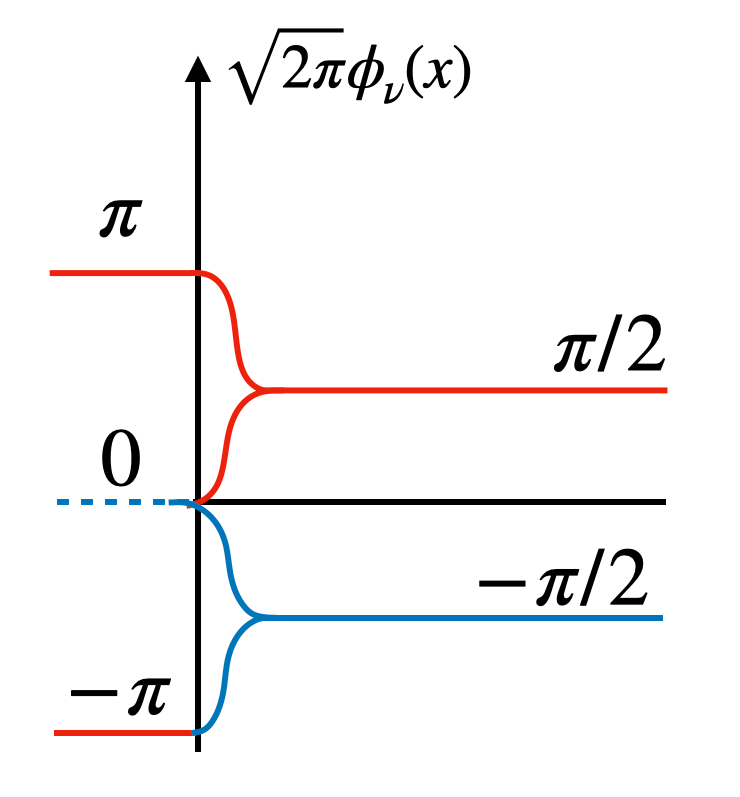}
\caption{}
\label{metal_pic_top}
\end{subfigure}
\caption{Bosonic ground state of gapped sectors of the gapless phases.  a) topologically trivial case with no edge modes.  It represents a trivial Mott insulator for $\nu=c$ and a trivial Luther-Emery liquid for $\nu=s$ b) topologically non-trivial state,  that has fourfold degenerate edge states.  For $\nu=c$ it corresponds to the edge states in topological Mott insulator and the $\nu=s$ case describes edge modes in topological Luther-Emery liquid.  }
\label{metal}
\end{figure}

\begin{figure*}[t]
    \centering
    
    \begin{minipage}{0.48\textwidth}
        \centering
        \includegraphics[scale=0.12]{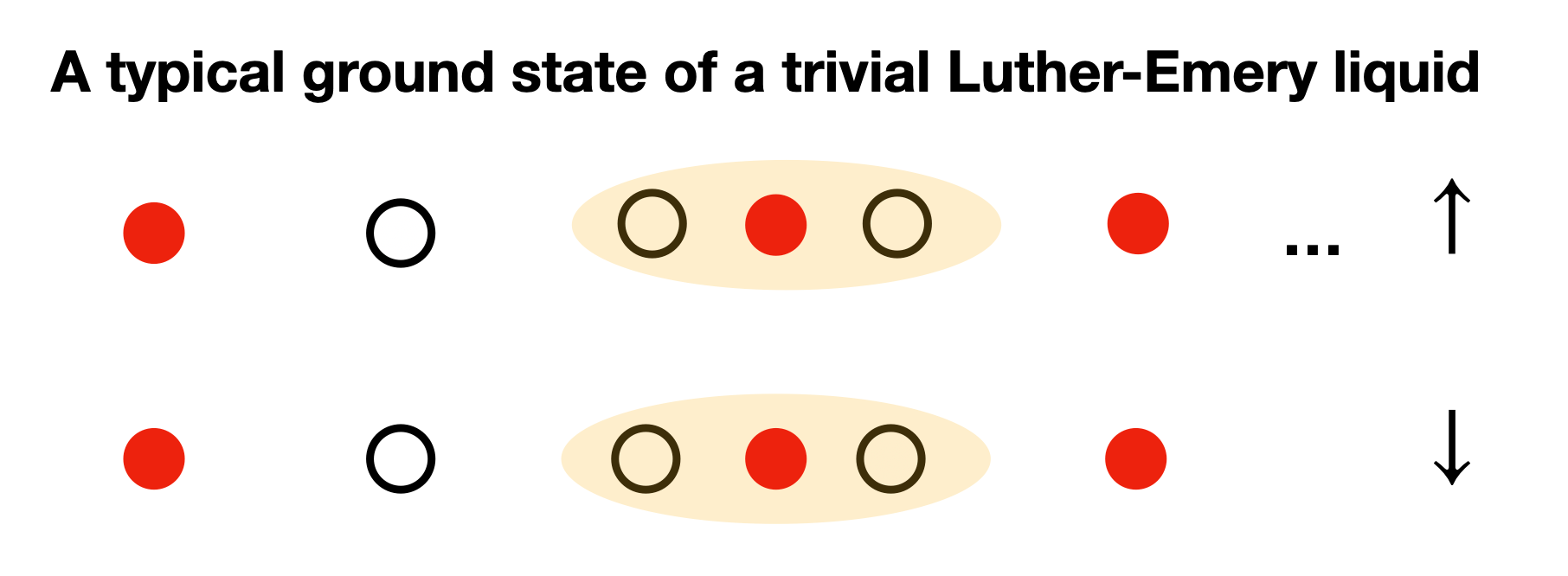}
        \caption{A snapshot of the ground state in a trivial Luther-Emery liquid.  Red dots schematically illustrate positions of electrons and empty black dots correspond to empty sites.   }
        \label{triv_LuE:GS}
    \end{minipage}
    \begin{minipage}{0.48\textwidth}
        \centering
        \includegraphics[scale=0.10]{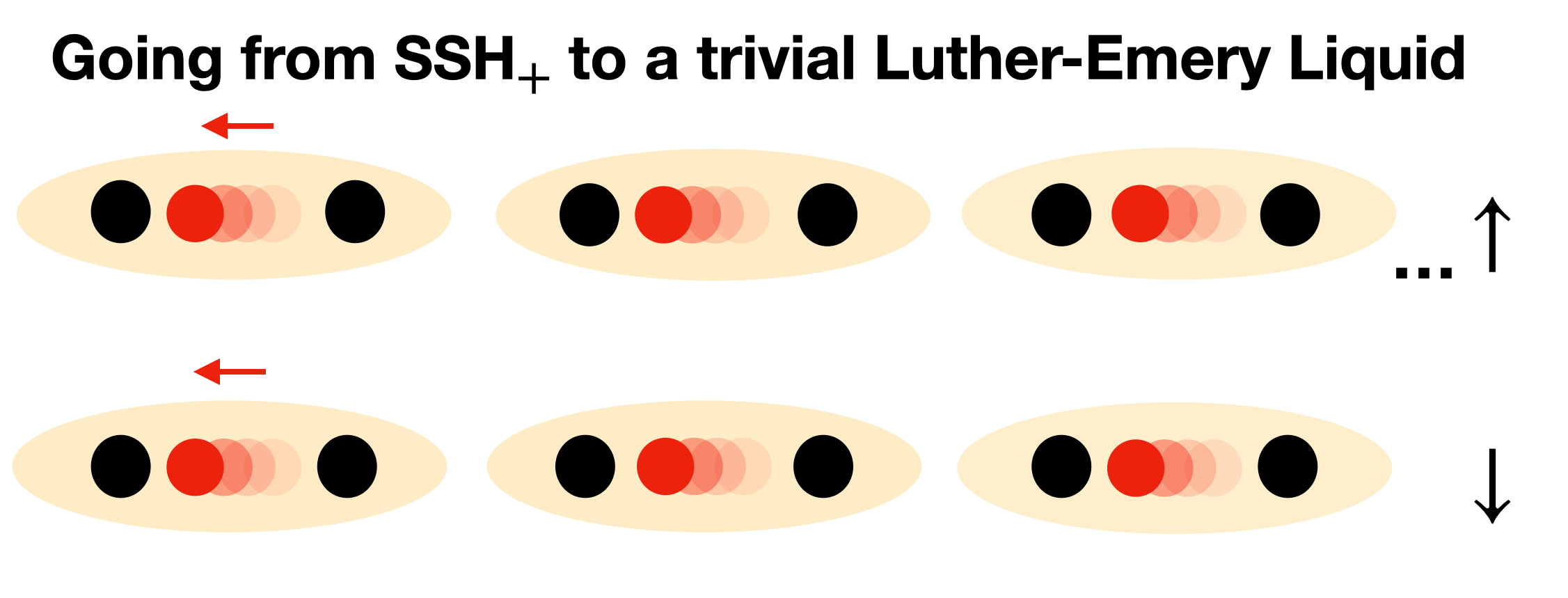}
        \caption{Fluctuations of charges in the ground state of a trivial Luther-Emery liquid.  The ground state can be thought as $\text{SSH}_+$ state with fluctuating charge positions.  }
        \label{triv_LuE:excited}
    \end{minipage}
    \hfill

\end{figure*}
\begin{figure*}[t]
    \centering
        \begin{minipage}{0.48\textwidth}
        \centering
        \includegraphics[scale=0.12]{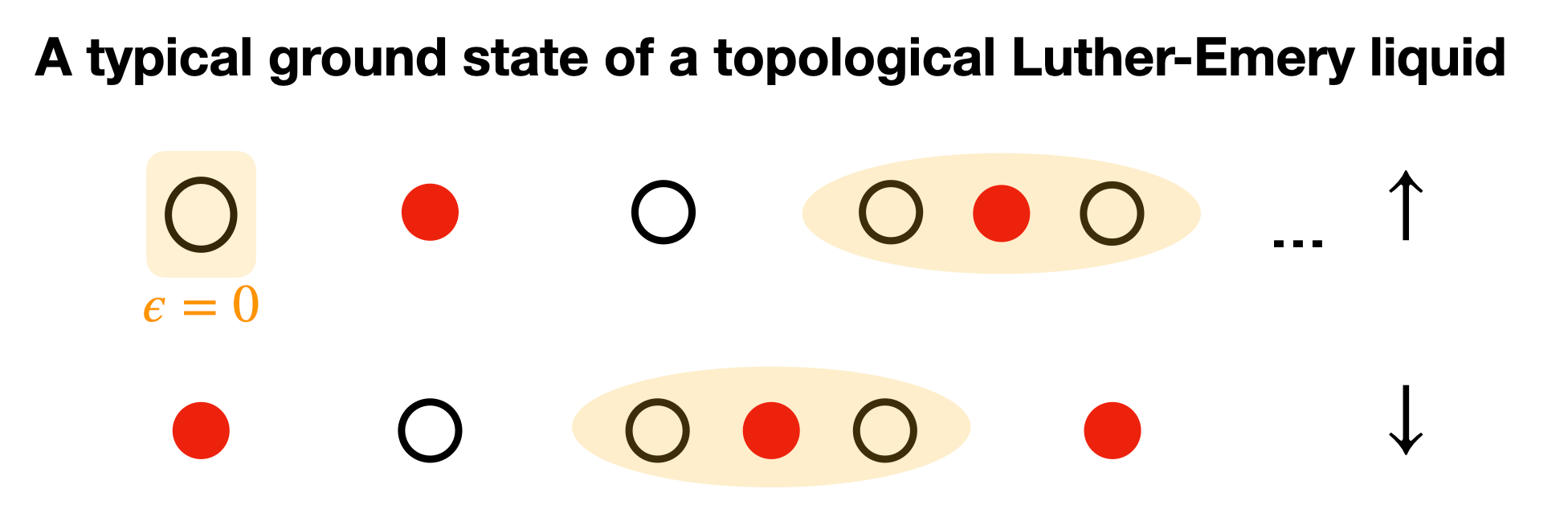}
        \caption{A snapshot of a ground state in a topologically non-trivial Luther-Emery liquid.  Note that there is a decoupled mode that is localized at the boundary.  }
        \label{top_LuE:GS}
    \end{minipage}
    \begin{minipage}{0.48\textwidth}
        \centering
        \includegraphics[scale=0.10]{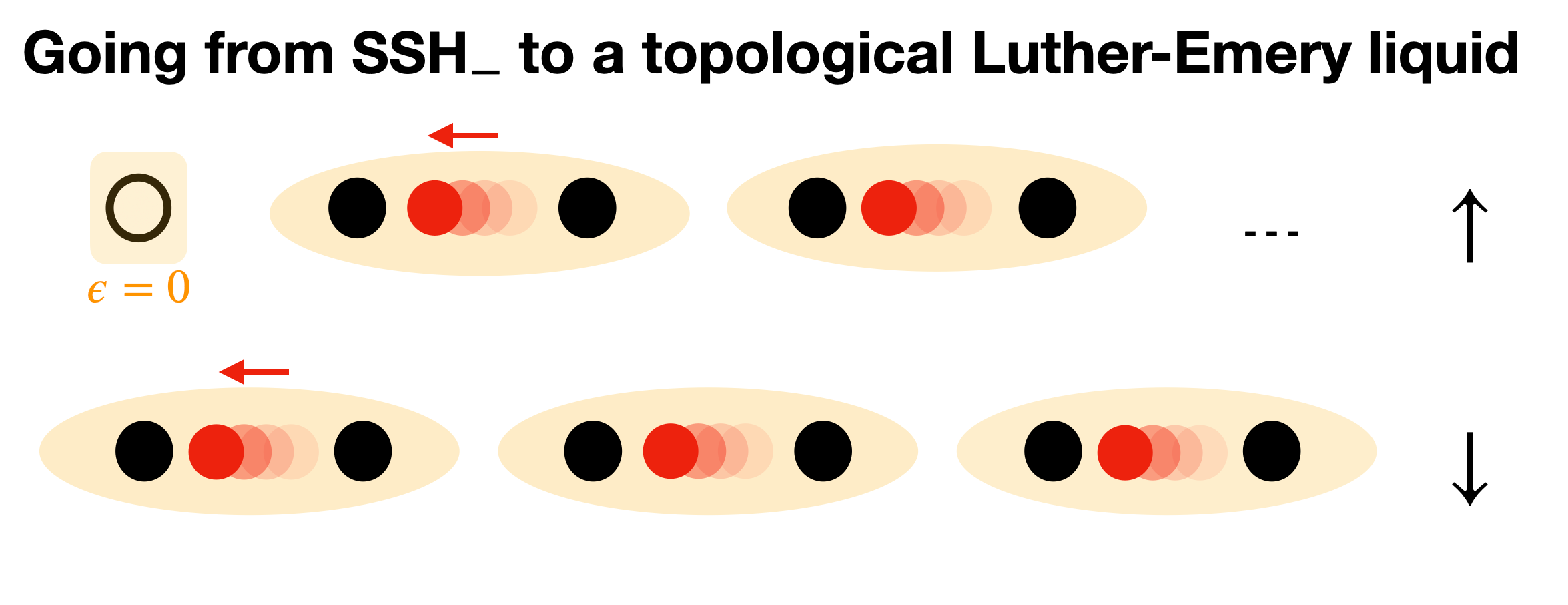}
        \caption{Fluctuating ground state in a topological Luther-Emery liquid.  The edge mode remains decoupled. }
        \label{top_LuE:excited}
    \end{minipage}
    \hfill
\end{figure*}

\subsubsection{General properties of Luther-Emery liquids}
Let us first focus on the line $g_c=0$ where the model is metallic.  The phases that occur on that line correspond to a topological Luther-Emery liquid for $g_s>0$ and a trivial Luther-Emery liquid \cite{Luther1974}  for $g_s<0$.  Let us discuss some of their general properties first.

First of all,  even though both Luther-Emery phases are metallic, the single-particle tunneling is supressed due to the spin gap.  However,   the two electron tunneling is allowed and costs no energy in a thermodynamic limit.  The two particles should have an opposite spin to be consistent with the spin order.  The precise spin state of the two electrons - singlet or triplet,  depends on a phase.  We will discuss the  details below.  
%
%
Secondly,   while in the half-filled case Luther-Emery phases occur  on the line $g_{c}=0$,  these phases can be extended to the region $g_{c} \neq 0$ if we dope the model it to shift it away from half-filling.
Doping changes the chemical potential that makes the charge sector gapless at the energies smaller than $v_F \cdot \delta k$,  where $\delta k$ is the deviation of Fermi momentum from $\pi/2a$.

\subsubsection{Trivial Luther-Emery liquid}

A trivial Luther-Emery liquid represents a phase boundary between the gapped phases $\text{CDW}$ and $\text{SSH}_+$,  so the ground state is represented by a superposition of states that combine bond- and charge-density order as illustrated in Fig.  \ref{triv_LuE:GS}.  This static picture represents only a snapshot of one possible ground state, since in the true ground state the electron positions fluctuate.  One way to visualize this dynamical behavior is start with the gapped phase,  for example,  $\text{SSH}_+$ where the ground state is dimerized.  As we approach the gapless line $g_c=0$,  the positions of electrons start to fluctuate between the atoms as shown in Fig.  \ref{triv_LuE:excited},  so we get a state that alternates between $\text{CDW}$ and $\text{SSH}_+$ orders.  

Just like in the original Luther-Emery model or in attractive Hubbard model,  the excited states can be thought as singlet pairs of electrons that move freely through the chain.   Because of that,  one can tunnel a singlet into the Luther-Emery liquid without extra energy cost.  In particular,  the operator that creates two electrons in a singlet state is proportional to $ \cos[\sqrt{2\pi} \phi_s] \exp[i\sqrt{2\pi}\theta_c]$.  If we compute the expectation value of this operator over the spin degrees of freedom,  taking into account that in a trivial Luther-Emery liquid $\sqrt{2\pi} \phi_s = 0 \mod \pi$,  we obtain that the singlet tunneling amplitude is nonzero.  We also see that the phase is topologically trivial as both  $\text{CDW}$ and $\text{SSH}_+$ phases do not have edge modes,  and therefore the gapless state between them is also non-topological.   We see that also from the bosonic language in Fig.   \ref{metal_pic_triv},  that shows that there are no edge solitons in a finite system.    
Our phase has some differences compared to the original Luther-Emery liquid,  in particular the dominant charge fluctuations are charge-density wave and $\text{SSH}_+$ in our phase,  while in the original phase they are superconducting ($K_c>1$ in case of attractive interactions).  

\subsubsection{Topological Luther-Emery liquid}
The topological Luther-Emery liquid occurs at the boundary between $\text{SDW}$ and $\text{SSH}_-$ phases.  The ground state is illustrated in Fig.  \ref{top_LuE:GS}.  It can be represented as a superposition of the states that have antiferromagnetic spin order and charge localised on sites or on bonds.  The charge fluctuations in this state are illustrated in Fig.  \ref{top_LuE:excited},   they correspond to electrons with the opposite spin fluctuating in the same directions.  
 Similar to the trivial Luther-Emery liquid,  one can tunnel two electrons into the bulk without extra energy cost,  however the two particles should be in a triplet state.  
The triplet creation operator with $S_z=0$ is proportional to $ \epsilon \cdot \partial_x \phi_c \sin[\sqrt{2\pi} \phi_s] \exp[i\sqrt{2\pi}\theta_c]$,  where $\epsilon$ - is the spatial separation between two electrons.   In the topological Luther-Emery phase $\sqrt{2\pi} \phi_s = \pi/2 \mod \pi$ and therefore the expectation value of this operator is non-zero.  

The topological Luther-Emery liquid features degenerate edge modes illustrated in Fig.  \ref{metal_pic_top} as bosonic edge solitons and in Fig.  \ref{top_LuE:GS} and Fig.  \ref{top_LuE:excited} within a fermionic picture.  
 Presence of such modes implies that it costs no extra energy to add a single electron to the edge,  as this would not affect the spin order in the bulk.  These edge states are quite peculiar and are physically different from the single-particle edge states in non-interacting insulators.  In particular,  when one adds an electron in a certain spin state at the edge,  for  the example shown in Fig.  \ref{top_LuE:GS} it would be a spin up particle,   the charge becomes redistributed along the chain (as there is no charge gap),  however the spin of this state is localized near the edge as there is a spin order in the bulk. 
Another possibility,  that also does not cost any extra energy,  is removing a spin-down particle from the edge,  that would cause similar charge fluctuations.   In fact,  the edge states can be represented as a superposition $\left(\alpha c^{\dagger}_{\sigma} +\beta c_{-\sigma}\right)|\text{vacuum} \rangle$,  where the creation and annihilation operators act on the edge.  Such edge states carry a fractionalized spin $s=\pm 1/4$ and were studied in different microscopic models that host the same topological Luther-Emery phase that we are focusing on.  They include one-dimensional wires with spin-orbit coupling \cite{Kainaris2015}, wires with spin-anisotropic interactions \cite{Kainaris2017,Verresen2021},  two interacting chiral modes at the edges of QSH planes \cite{Kainaris2017,Santos2016} and superconducting wires \cite{Keselman2015,Andrei2020}.  
In Ref.  \cite{Verresen2021} it was shown that such a topological liquid is protected by $\mathbb{Z}_4$ symmetry that is broken when one opens up a gap.  In their microscopic model the $\mathbb{Z}_4$ symmetry represented by a $\pi$ - rotation,  while in our work we associate the $\mathbb{Z}_4$ symmetry with real-space translations by half of the lattice constant.  Indeed,  we see from Fig.  \ref{top_LuE:GS} that shifting the ground state by $a/2$ produces another valid ground state.   This transformation is a subgroup of the translation symmetry that is preserved by the gapless state. 
It breaks down to $\mathbb{Z}_2$ when we open up a charge gap.  As we see from the illustrations in Fig.  \ref{phase_diagram},  in SDW and $\text{SSH}_-$ phases are symmetric with respect to the shift by $a$ that produces the state with the same energy,  and applying this shift twice brings us back to the original state.  

\begin{figure*}[t]
    \centering
    \begin{minipage}{0.48\textwidth}
        \centering
        \includegraphics[scale=0.13]{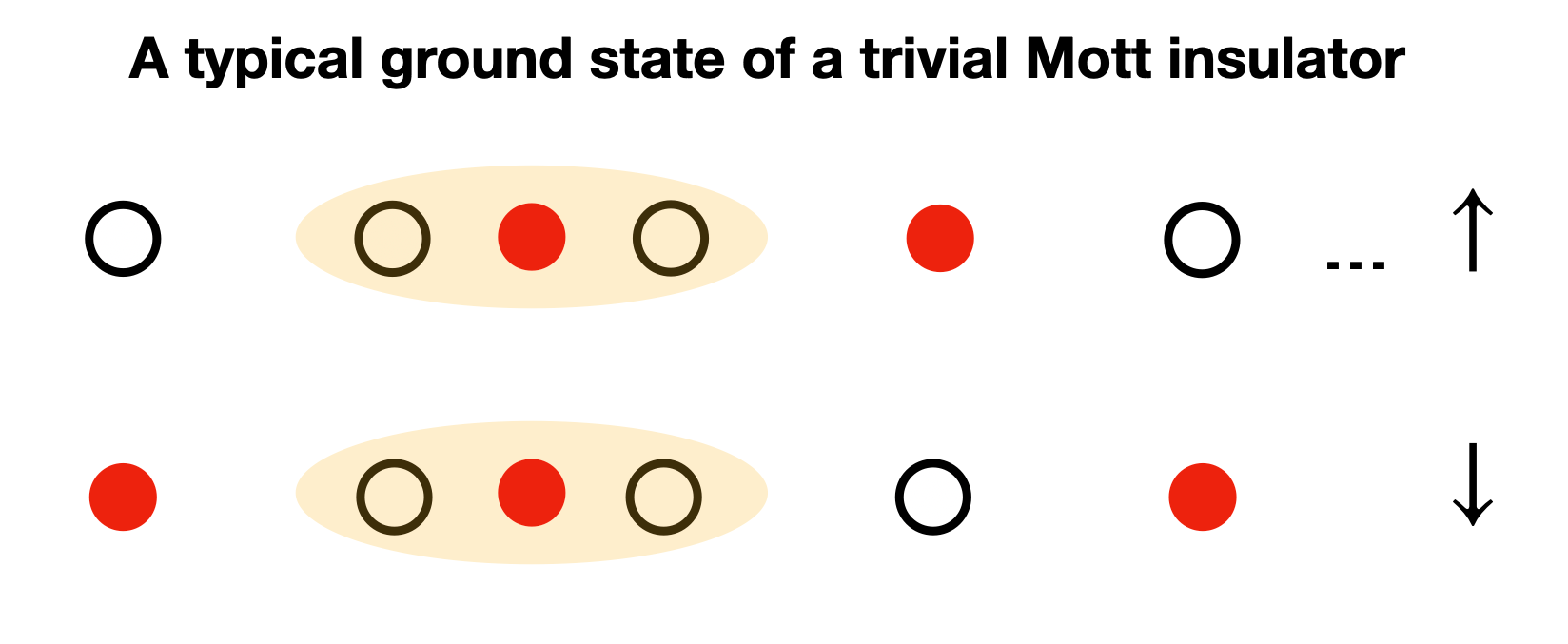}
        \caption{A snapshot of a ground state in a trivial Mott insulator.  }
        \label{triv_Mott:GS}
    \end{minipage}
    \hfill
    \begin{minipage}{0.48\textwidth}
        \centering
        \includegraphics[scale=0.13]{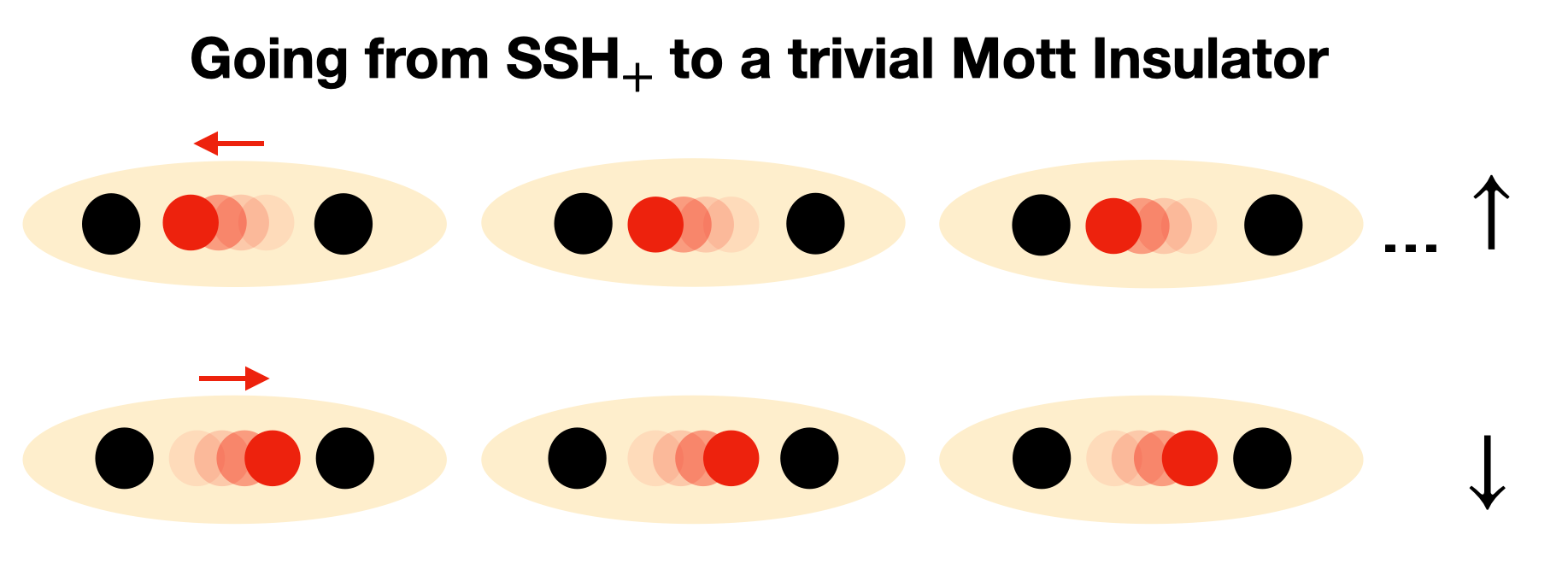}
        \caption{Illustration of the fluctuating charges in ground state of a trivial Mott insulator. }
        \label{triv_Mott:excited}
    \end{minipage}

\end{figure*}

\begin{figure*}[t]
    \centering
      
    \begin{minipage}{0.48\textwidth}
        \centering
        \includegraphics[scale=0.13]{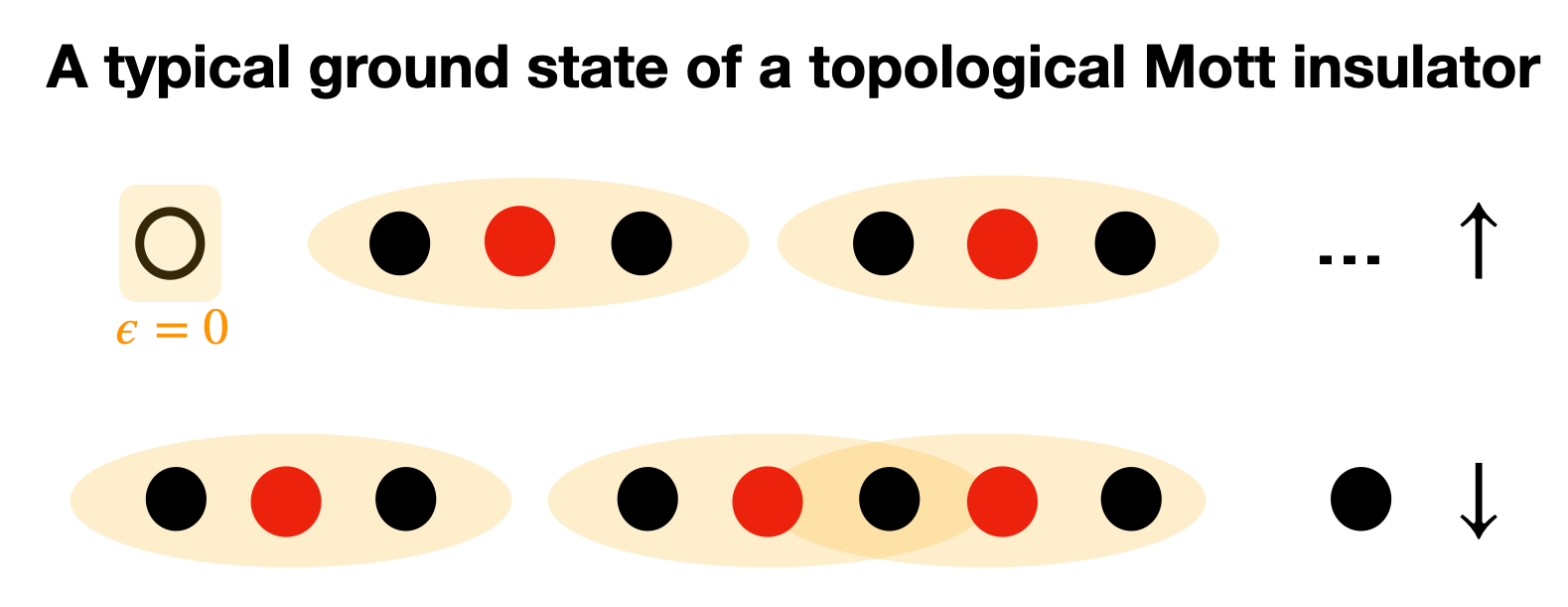}
        \caption{Ground state of a topological Mott insulator featuring a decoupled edge mode.}
        \label{top_Mott:GS}
    \end{minipage}
	\hfill
    \begin{minipage}{0.48\textwidth}
        \centering
        \includegraphics[scale=0.13]{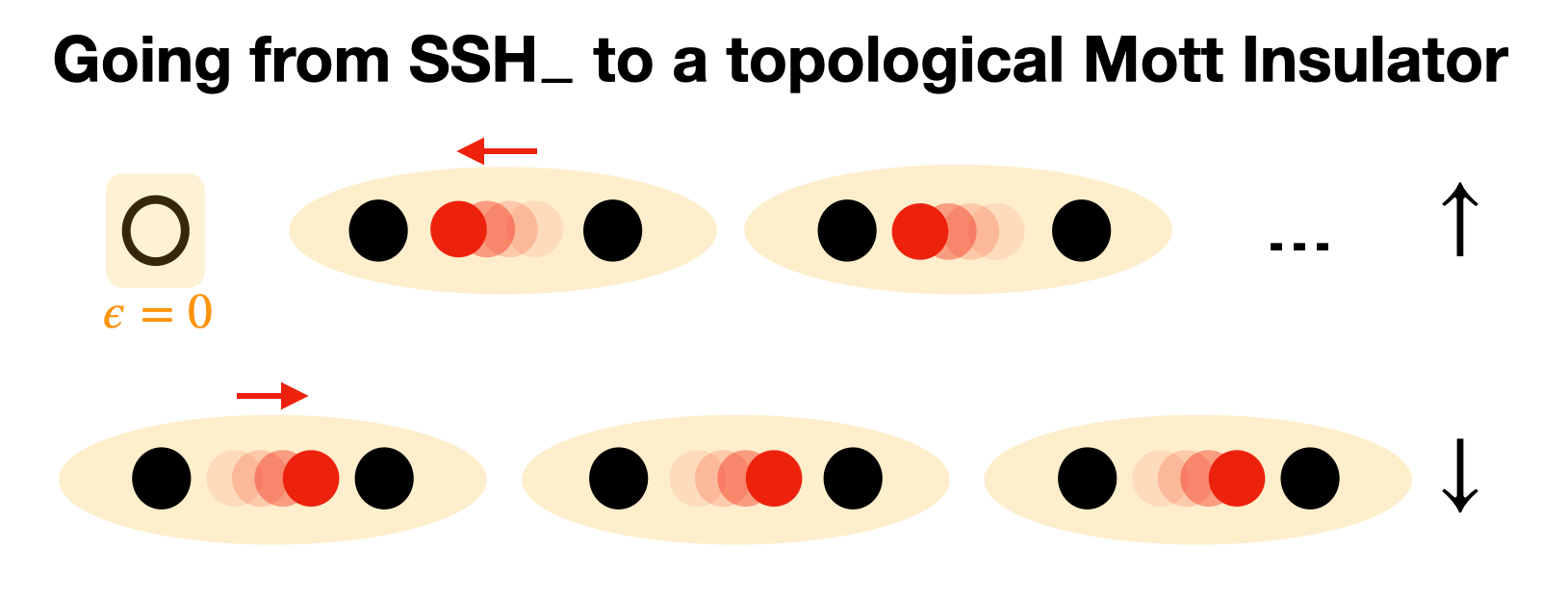}
        \caption{Ilustration of the fluctuating ground state of a topological Mott insulator.  The edge mode remains decoupled. }
        \label{top_Mott:excited}
    \end{minipage}

\end{figure*}

\subsection{Trivial and topological Mott insulators}
Now we focus on a line $g_s=0$,  where the spin sector is gapless and charge is gapped.  
The phase between $\text{SDW}$ and $\text{SSH}_+$ is equivalent to Mott insulator.  The phase boundary between $\text{SSH}_-$ and $\text{CDW}$ features degenerate edge modes and corresponds to a topological Mott insulator.   Similar to the Luther-Emery states,  the Mott insulators can be extended to the region $g_s \neq 0$
by adding the Zeeman term to the Hamiltonian.  The Zeeman term removes spin degeneracy of the spectrum and splits it for different spin components in $k$-space.   The splitting does not affect the  charge sector and total filling,  however the gap-opening term in spin sector becomes oscillating,  and thus can be neglected at the energies smaller than $v_F \cdot \delta k$.   Let us now discuss the Mott phases in more detail.  

\subsubsection{Trivial Mott insulator}
A trivial Mott insulator occurs at the phase boundary between SDW and $\text{SSH}_+$ phases.  Therefore,  the ground state of the Mott phase can be represented as a superposition of states that combine the two orders.  One example of a ground state is shown in Fig.  \ref{triv_Mott:GS}.  The fluctuations in this ground state are illustrated in Fig.  \ref{triv_Mott:excited}.  Similar to a trivial Luther-Emery liquid,  they can be thought as charges that fluctuate in the $\text{SSH}_+$ phase,  however in a Mott insulator particles with opposite spin move in the opposite directions,  because there is a charge gap.  As in a trivial Luther-Emery liquid,  there are no decoupled edge modes in a trivial Mott insulator.   

Note that this ground state looks different from a typical Mott insulator,  that represents the cartoon ground state of the repulsive Hubbard model at half-filling.  In the Hubbard model the charges are localized at the lattice sites,  while in our model they are allowed to fluctuate between the sites.   However,  in the Appendix \ref{app:Mott} we show that the two phases are adiabatically connected.

\subsubsection{Topological Mott insulator}
Now let us focus on topological Mott insulator that occurs at a phase boundary between the CDW phase and $\text{SSH}_-$.  The ground state of this phase in shown in Fig.  \ref{top_Mott:GS}.  Similar to the other gapless states,  the ground state of the topological Mott insulator is a superposition of states that combine orders of the neighbouring gapped phases.   The fluctuations in this state are illustrated in Fig.  \ref{top_Mott:excited} and correspond to particles with different spin fluctuating in the opposite directions,  similar to a trivial Mott insulator.  However,  there is always a decoupled state near the edge that represent a topological edge mode.  This state is characterized by a fractional charge $Q=\pm e/2$ which is localized near the edge,  while the spin degree of freedom is delocalized.   
We also expect that topological Mott insulator is protected by $\mathbb{Z}_4$ symmetry,   associated with translations by $a/2$, similar to the topological Luther-Emery phase.  This symmetry is reduced to $\mathbb{Z}_2$ when a gap is opened in the spin sector, leading to either the CDW or $\text{SSH}_-$ phase.
Note that topological Mott insulator phase was identified and studied numerically in Ref.  \cite{Yoshida2014} in an interacting spinful SSH model.  

\section{Connecting interacting and non-Interacting gapless phases}
\label{sec:connection}
 In this section we demonstrate that our topological gapless phases can be adiabatically deformed to non-interacting topological metals discussed in Section \ref{sec:nonint}.  

\subsection{Phase diagram of a generalized model}
\begin{figure}
\includegraphics[scale=0.15]{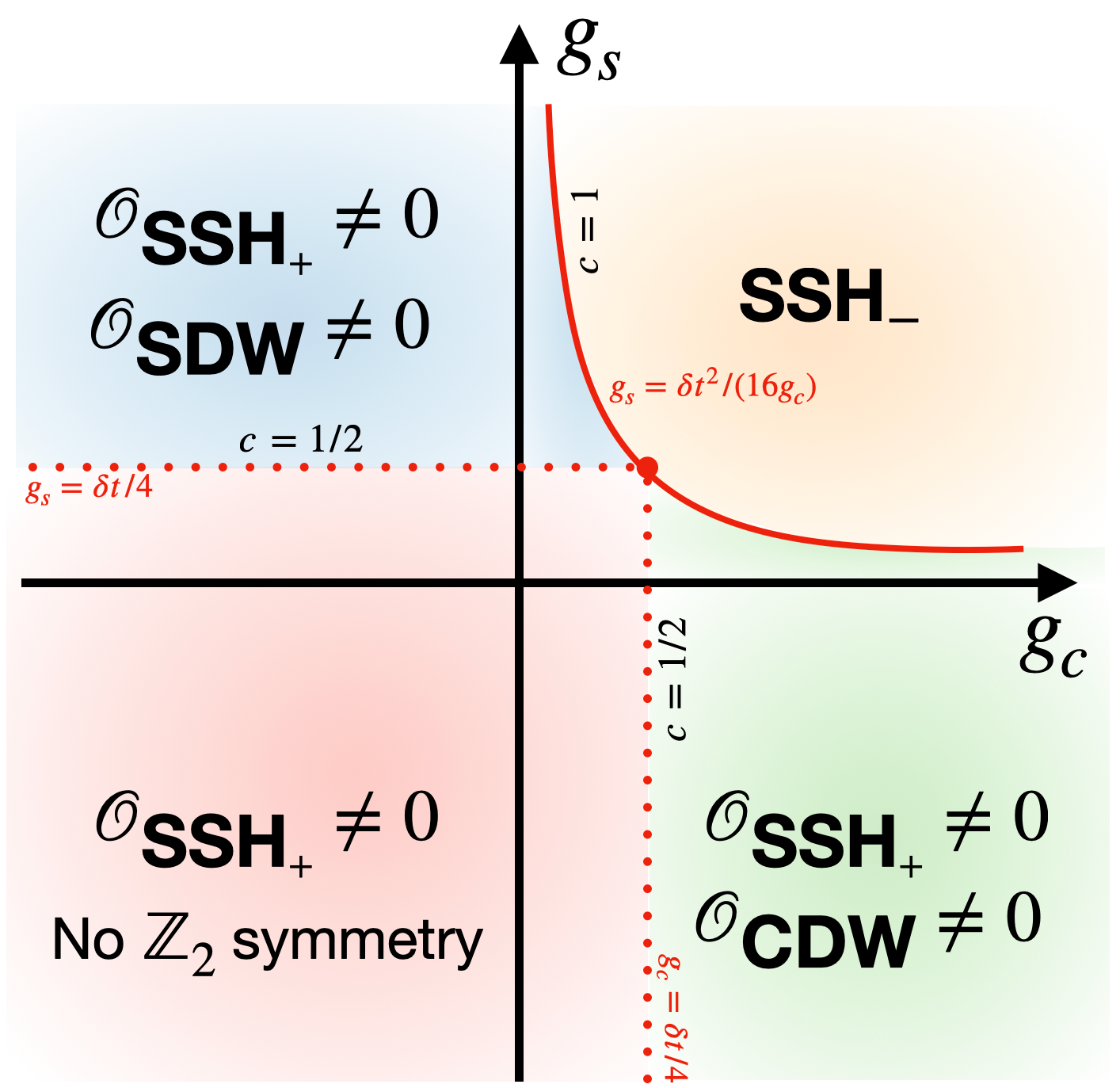}
\caption{Phase diagram for the model \eqref{gap_full_deltat} for $\delta t= \text{const}>0$.  The red dotted lines correspond to boundaries between the phase without $\mathbb{Z}_2$ symmetry to the mixed phases with spontaneously broken $\mathbb{Z}_2$ symmetry.   
At these lines the model undergoes phase transition of Ising type in one of the sectors: at $g_c=\delta t/4$  the charge sector undergoes the phase transition and the line $g_s=\delta t/4$ corresponds to the spin degrees of freedom are gapless.  The red solid line between the phases with spontaneously broken $\mathbb{Z}_2$ symmetry corresponds to the second order phase transition.  The gapless excitations along this line are described by Luttinger Liquid.   
 }
\label{phase_deltat}
\end{figure}

To construct the adiabatic path between interacting and non-interacting phases,  we generalize our model \eqref{full_model_main} and add the following single-particle term: 
\begin{align}
\label{Vprime}
V_{\nu=2} = \frac{\delta t}{(\pi a)^2} \cos[\sqrt{2\pi} \phi_c] \cos[\sqrt{2\pi} \phi_s].  
\end{align}
The full gap-opening part of the bosonic Hamiltonian including \eqref{Vprime} and \eqref{gap_terms_main}  becomes: 
\begin{align}
\label{gap_full_deltat}
& H_{\text{gap}}=\frac{g_c}{(\pi a)^2} \cos[\sqrt{8\pi} \phi_c]+\frac{g_s}{(\pi a)^2} \cos[\sqrt{8\pi} \phi_s] + \nonumber \\ 
& + \frac{\delta t}{\pi a} \cos[\sqrt{2\pi} \phi_c] \cos[\sqrt{2\pi} \phi_s]. 
\end{align}
In fermionic language the $\delta t$ term corresponds to the Hamiltonian that describes two identical non-interacting SSH chains.  For $\delta t>0$ it corresponds to a phase with winding number $\nu=2$.  This term also coincides with the expression for the order parameter in $\text{SSH}_+$ phase.  The phase diagram of the model \eqref{gap_full_deltat} is shown in Fig.  \ref{phase_deltat} for some fixed $\delta t>0$.  

The $\delta t$ term breaks $\mathbb{Z}_2$ symmetry in $\text{SSH}_+$ phase,  because it is compatible with only one of two bulk minima in that phase for a fixed sign of $\delta t$.   However,  $\mathbb{Z}_2$ symmetry is still present in all remaining phases.  In spin-density and charge-density wave phases this term induces the  $\text{SSH}_+$ order and it shifts the phase boundaries of $\text{SSH}_-$ phase.   
Let us now focus on the gapless boundaries between the gapped phases to see what changes in their physical properties when we add the additional term.   

First, the two lines indicated by red dots in Fig. \ref{phase_deltat} correspond, in the original phase diagram, to the trivial Mott insulator and the trivial Luther–Emery liquid.  In the new phase diagram, these lines \footnote{Strictly speaking,  the phase boundary is determined by the condition $\Delta_c \simeq \Delta_{\delta t}$,  where $\Delta_c$ and $\Delta_{\delta t}$ are the energy gaps associated with the corresponding cosine terms.  Taking into account that $\Delta_c \propto g_c^{1/(2-2K_c)}$ and $\Delta_{\delta t} \propto \delta t^{2/(4-K_c-K_s)}$ we see that we can replace the gap expressions by the coupling constants themselves as we do in the main text when $K_{s,c} =1/2$ that happens when interactions are strong.  Similar arguments apply to the other phase boundaries. }  represent transitions of the Ising universality class with $c=1/2$, occurring in either the charge or spin sector \cite{gogolin2004bosonization}.  We expect this type of phase transition to occur when moving from a phase without $\mathbb{Z}_2$ symmetry to the phases with spontaneously broken  $\mathbb{Z}_2$ symmetry.   
Note that along these lines, one sector remains gapped because the minima of the $\delta t$ term are compatible with those in the spin sector of the SDW phase and in the charge sector of the CDW phase.

\subsection{Adiabatic path between interacting gapless phases and non-interacting metals }

Let us now look at the critical phases between the $\text{SSH}_-$ phase and all the remaining phases.   First of all we see that instead of two separate phase boundaries, one for former topological Mott insulator at $g_s=0$ and one for topological Luther-Emery liquid at $g_c=0$,  we now have a single critical line given by $g_s =\delta t^2/16g_c$.   We find that along this line the model is gapless and the gapless sector described by a Luttinger Liquid.   We find that the gapless degrees of freedom smoothly interpolate between charge and spin modes in the limits $g_s \rightarrow \infty$ and $g_c \rightarrow \infty$,  respectively,  and become spin-up or spin-down modes at the special critical point $g_c=g_s=\delta t/4$.  At this point the model can be smoothly connected to a non-interacting metal discussed in Section \ref{sec:nonint}.

Let us focus on these two simple limits first.  In the limit $g_{c,s} \rightarrow \infty$ we reproduce the phase boundaries of the original model,  where one of the fields -  charge or spin becomes gapless.  To better understand what happens at the critical point $g_c=g_s=\delta t/4$ let us examine the ground state of the bosonic Hamiltonian \eqref{gap_full_deltat} in this special phase:  
\begin{align}
\label{H_gapped_point}
& H_{\text{gap}}  \underset{\substack{\text{critical} \\ \text{point}}}{=} \frac{\delta t}{2(\pi a)^2} \left(\cos[\sqrt{4\pi} \phi_{\uparrow}] +\cos[\sqrt{4\pi} \phi_{\downarrow}] +  \right.   \nonumber \\
& + \left.\cos[\sqrt{4\pi} \phi_{\uparrow}] \cdot \cos[\sqrt{4\pi} \phi_{\downarrow}]  \right), 
\end{align}
where we used bosonization notations from Appendix \ref{bosonization_conventions}.
From semiclassical analysis,  one set of minima corresponds to $\sqrt{4\pi} \phi_{\uparrow} =\pi \mod 2\pi$.  Note that at these minima the Hamiltonian \eqref{H_gapped_point} is independent of  $\phi_{\downarrow}$ field.  Therefore,  the spin down degree of freedom is gapless.  Another possible set of minima is given by $\sqrt{4\pi} \phi_{\downarrow} =\pi \mod 2\pi$.  In this state spin up degrees of freedom are gapless.  It is now easy to see that the Hamiltonian (\ref{H_gapped_point}) can be adiabatically deformed to a non-interacting metal.  One can show that by adding another single-particle term: 
\begin{align}
V_{\nu=1}= \pm \frac{\delta t}{2(\pi a)^2} \left( \cos[\sqrt{4\pi}\phi_{\uparrow}] - \cos[\sqrt{4\pi}\phi_{\uparrow}] \right)
\end{align}
This term describes two decoupled SSH chains in the opposite phases with the total winding number $\nu=1$.   The sign in front of this term determines which one of the two chains is topological.   Depending on that,  the minima of this term are compatible with either $\sqrt{4\pi} \phi_{\uparrow} =\pi \mod 2\pi$ minima of the model \eqref{H_gapped_point} where  the spin down degree of freedom is gapless,  or with $\sqrt{4\pi} \phi_{\downarrow} =\pi \mod 2\pi$ where the spin up field is gapless.   After adding this term,  we can adiabatically switch off interactions in \eqref{H_gapped_point}.  The remaining non-interacting bosonic model corresponds to the topological metal that we discussed in Section \ref{sec:nonint}.  This is very similar to writing a mean-field for the gapped phases,  where we deform a many-body symmetry-breaking state into a single-particle model by adding non-interacting terms that are compatible with one of the degenerate ground states.  

\begin{figure}[t]
\raggedright
\includegraphics[scale=0.55]{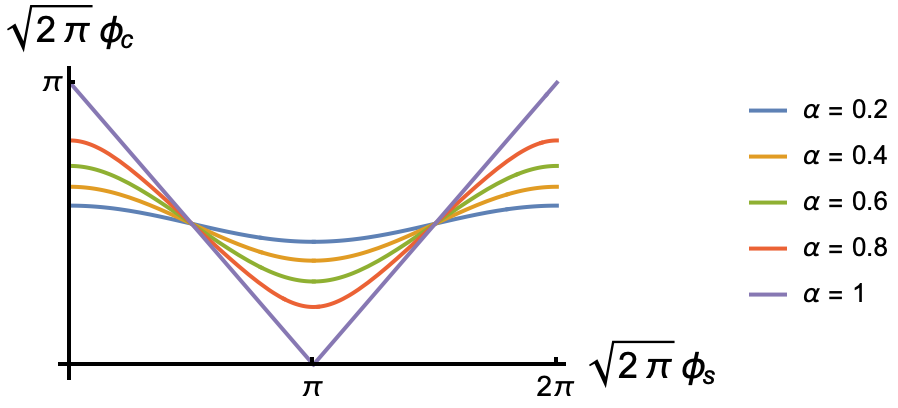}
\caption{The relation between spin and charge fields in the minima of the Hamiltonian at the critical line $g_s = \delta t^2/(16g_c)$.  Excitations along these lines are gapless.  Here $\alpha= \delta t/4g_c$.  The case $\alpha=1$ corresponds to a special point where $g_c=g_s=\delta t/4$,  where the gapless excitations correspond to spin up or spin down degrees of freedom.  In this point a model can be deformed to a non-interacting topological metal.  
 }
\label{gapless_line_phi}
\end{figure}

Away from the special point $g_c=g_s$ on a critical line the gapless degrees of freedom are \textit{nonlinear} combinations of the charge and spin fields.  The positions of bosonic minima of the Hamiltonian \eqref{gap_full_deltat} is given by  $\sqrt{2\pi}\phi_c = \pi  \pm  \arccos \left((\delta t/4g_c) \cos[ \sqrt{2\pi}\phi_s]\right) \mod 2\pi$.  Along these lines the excitations are gapless.   One of these gapless lines (with the minus sign) is illustrated in Fig.  \ref{gapless_line_phi}.  We see that away from the special point $g_c=g_s$ this relation becomes non-linear.  We discuss the spectrum of gapped and gapless excitations in a general case in Appendix \ref{gapless}.  

\section{Discussion and conclusion}
We studied topology of gapless states in a fermionic model with spin-charge separation.  In particular,  we focused on the phase boundaries between the gapped states with spontaneously broken $\mathbb{Z}_2$ symmetry.   

Along those boundaries one of the sectors becomes gapless,  while the other remains gapped.  We studied the edge states in this model using bosonization,  following the ideas from our previous works \cite{Matveeva2025,Matveeva2024},  and identified two topologically non-trivial gapless states,  a topological Luther-Emery liquid and a topological Mott insulator.  These phases occur at the boundary between a topological insulator and trivial insulators.  Topological Luther-Emery phase is characterized by gapless charge excitations,  while spin sector is gapped and features edge modes with fractionalized spin $s_z=\pm 1/4$.  We find that similar to a regular Luther-Emery liquid,  one can tunnel two electrons into such state without extra energy cost in a thermodynamic limit.  Such a pair should be in a triplet state in a topological liquid,  while in a trivial liquid one can tunnel singlets into the bulk.  A topological Luther-Emery phase was studied before in many microscopic models \cite{Keselman2015, Kainaris2015,Santos2016,Kainaris2017,Andrei2020,Verresen2021}.  
A topological Mott insulator is characterized by gapless spin excitations, while the charge sector is gapped and hosts edge modes carrying fractional charge $Q=\pm e/2$.  

We also discussed symmetries that protect the topological phases.  In particular,  we use the results obtained in Ref.  \cite{Verresen2021},  where the authors focused on topological Luther-Emery liquid in Ising-Hubbard model and proved that it is protected by a $\pi$ rotation around $x$ axes which represents $\mathbb{Z}_4$ symmetry,  when it acts on many-body Hilbert space.  In our model we associate this symmetry with translations by half of the lattice constant.  This symmetry is preserved in the gapless phase,  and gets reduced to $\mathbb{Z}_2$ in the gapped states where the charges are localized  either at the lattice sites or in the bonds between the lattice sites.  
The same effective bosonic model was analyzed in Ref.~\cite{Montorsi2017},  and our findings for the topology of the gapless states are in agreement with theirs.

Next we showed that in a more general model our interacting gapless states can be  adiabatically deformed to non-interacting topological metal that occurs at the boundary between the insulators with winding number $\nu=2$ and $\nu=1$.  In between these two phases we find a gapless state where the excitations are some non-linear combinations of spin and charge.  It would be interesting to study the  edge modes in such an intermediate phase and see how they interpolate between the single-particle edge states in a non-interacting metal and many-body edge modes in the interacting model.  

\section*{Acknowledgments}
We thank Natan Andrei and Elio K\"onig for the useful discussions.  
P.M.  acknowledges support from the US National Science Foundation (NSF) Grant Number
2201516 under the Accelnet program of Office of International Science and Engineering
(OISE). 

\onecolumngrid
\appendix
\section*{Appendix} 
\section{Stability of the edge modes in a non-interacting topological metal}
\label{app:stability}
In Section \ref{sec:nonint} we described the simplest non-interacting topological metal in a model of two decoupled chains.  It is worth discussing the topological stability of this phase,  in particular,  whether the edge modes survive adding a weak inter-chain coupling between them.  Consider the following coupling term that we add to the Hamiltonian \eqref{SSH_chains}: 
\begin{align}
\label{app:coupling_term}
H_{\perp} = g \sum_{n} \left(c_{A,1,n}^{\dagger} c_{B,2,n} + c_{A,2,n}^{\dagger} c_{B,1,n} + c^{\dagger}_{B,1,n} c_{A,2,n+1} +  c^{\dagger}_{B,2,n} c_{A,1,n+1} + \text{h.c.}  \right).    
\end{align}
This coupling is not dimerized and does not open a gap in the spectrum.  Now let us look for the edge state solutions of the full Hamiltonian that includes \eqref{SSH_chains} and \eqref{app:coupling_term}.   The zero-energy states satisfy the following equations of motion: 
\begin{align}
\label{app:equations_stability}
\begin{cases}
v_1 c_{A,1,n+1} + w_1 c_{A,1,n} +g(c_{A,2,n} + c_{A,2,n+1}) = 0 \\
v_2 c_{A,2,n+1} + w_2 c_{A,2,n} +g(c_{A,1,n} + c_{A,1,n+1}) = 0 \\
v_1 c_{B,1,n-1} + w_1 c_{B,1,n} + g(c_{B,2,n} + c_{B,2,n-1}) = 0 \\
v_2 c_{B,2,n-1} + w_2 c_{B,2,n} + g(c_{B,1,n} + c_{B,1,n-1}) = 0.
\end{cases}
\end{align}
In a half-infinite chain,  the wavefunction for a left-edge zero mode lives entirely on the A sublattice \cite{Matveeva2023},  so we set $c_{B,1,n} = c_{B,2,n} = 0$.
The remaining equations reduce to: 
\begin{align}
\label{app:equations_stability_zero}
v_1 c_{A,1,n+1} + w_1 c_{A,1,n} +g(c_{A,2,n} + c_{A,2,n+1}) = 0 \\
v_2 c_{A,2,n+1} + w_2 c_{A,2,n} +g(c_{A,1,n} + c_{A,1,n+1}) = 0. 
\end{align}
Next we look for solutions of the form: 
\begin{align}
\label{app:edge_solutions_form}
c_{A,1,n} = \alpha_1 \lambda^n,  \hspace{1cm} c_{A,2,n} = \alpha_2 \lambda^n
\end{align}
This leads to the following set of equations: 
\begin{align}
\begin{pmatrix}
w_1 + v_1 \lambda & g(1+\lambda) \\
g(1+\lambda) & w_2 + v_2 \lambda
\end{pmatrix}
\begin{pmatrix}
\alpha_1 \\
\alpha_2
\end{pmatrix}
= 0.
\end{align}
A nontrivial solution exists only if the determinant vanishes,  that gives the following roots when one of the chains is critical,  e.g.  $w_2=v_2$: 
\begin{align}
\lambda_1 = -1,  \hspace{1cm} \lambda_2 = -\frac{v_2w_1-g^2}{v_2v_1-g^2}.
\end{align}
A localized edge state exists whenever $|\lambda_2|<1$.  The critical value of the inter-chain coupling $g_c$ is determined by $\lambda_2 \equiv 1$ that gives the following $g_c$:  
\begin{align}
\label{app: gc}
g_c = \pm \sqrt{\frac{v_2(w_1+v_1)}{2}}.
\end{align}
Therefore the edge states are stable for $|g|<|g_c|$. 

\section{Bosonisation conventions}\label{bosonization_conventions}
\label{bosonization_spinful}
In the low energy limit,  the fermionic operator for a given spin is decomposed to right- and left-moving parts as follows: 
\begin{align}
c_{j,\sigma} \rightarrow \Psi_{\sigma}(x) = e^{ik_Fx} R_{\sigma}(x) + e^{-ik_Fx} L_{\sigma}(x)
\end{align}
The mapping between the chiral fermionic operators and bosonic fields is given by: 
\begin{align}
\label{RL_bosonizedS}
R_{\sigma}(x) = \frac{\kappa_{\sigma}}{\sqrt{2\pi a}} e^{i\sqrt{4\pi} \phi_{R\sigma}},  \hspace{0.3cm} L_{\sigma}(x) = \frac{\kappa_{\sigma}}{\sqrt{2\pi a}} e^{-i\sqrt{4\pi} \phi_{L\sigma}}. 
\end{align}
The Klein factors $\kappa_{\sigma}$ are Hermitian operators and they satisfy the algebra $\{\kappa_{\sigma}, \kappa_{\sigma'}\}= 2\delta_{\sigma \sigma'}$.  They ensure the correct anticommutation relations between  fermionic operators with different chain indexes.  The specific representation satisfying this algebra can be chosen such as $\kappa_{\uparrow}\kappa_{\downarrow} = -\kappa_{\downarrow}\kappa_{\uparrow}=i$.   The commutation relations between the bosonic fields are chosen such that fields with different chain index commute: 
 \begin{align}
\label{RLboson_commutators_spin}
[\phi_{R\sigma}(x), \phi_{L\sigma'}(y)] = \frac{i}{4} \delta_{\sigma,\sigma'}\nonumber \\ 
[\phi_{\eta, \sigma} (x), \phi_{\eta'\sigma'}(y)]=  \frac{i}{4} \eta \delta_{\eta, \eta'} \delta_{\sigma,\sigma'} \text{sign}(x-y),
\end{align}
here the indices $\sigma$ and $\sigma'$ denote the chain degree of freedom.  
It is useful to introduce the charge and spin bosonic fields: 
\begin{align}
\label{charge_spin_bosons}
\phi_c = \frac{\phi_{\uparrow}+\phi_{\downarrow}}{\sqrt{2}}, \hspace{0.3cm} \phi_s = \frac{\phi_{\uparrow}-\phi_{\downarrow}}{\sqrt{2}}.
\end{align}
One can use those fields to express the non-oscillatory part of charge and spin densities:  
\begin{align}
\rho_c(x) = \sum_{\sigma} \Psi^{\dagger}_{\sigma}(x) \Psi_{\sigma}(x)= \sqrt{\frac{2}{\pi}} \partial_x\phi_c, \nonumber\\ 
\rho_s (x)= \frac{1}{2} \sum_{\sigma\sigma'} \Psi_{\sigma}^{\dagger}(x) (\sigma_z)_{\sigma\sigma'}  \Psi_{\sigma'}(x)= \frac{1}{\sqrt{2\pi}} \partial_x\phi_s.  
\end{align}

\section{Connection between a regular Mott insulator and a general Mott insulator}
\label{app:Mott}
\begin{figure}[H]
\centering
\includegraphics[scale=0.15]{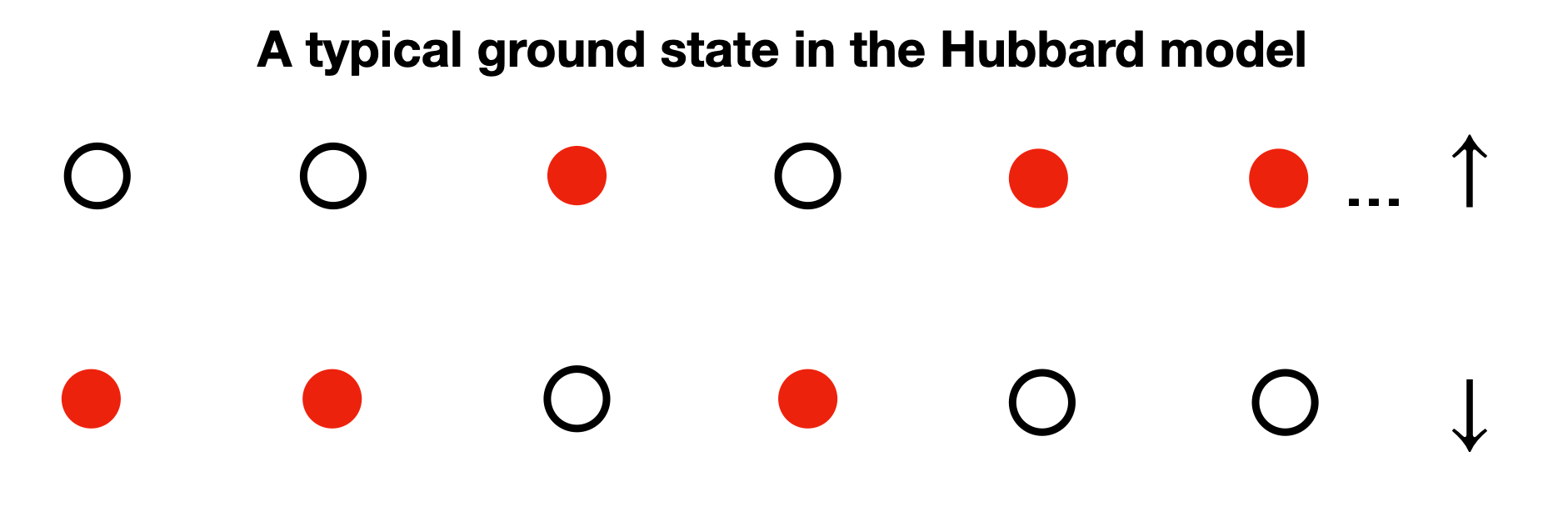}
\caption{A typical ground state of the Mott insulator phase in repulsive Hubbard model.  Red dots show filled states and empty dots show empty states. }
\label{MottInsulator_Hubb}
\end{figure}

Here we discuss the relation between a regular Mott insulator,  that occurs in a repulsive Hubbard model at half-filling,  and a trivial Mott insulator phase that we are focusing on.  In particular,  let us compare their ground states.  In a regular Mott insulator particles are localized on sites and spins are disordered.  The snapshot of one of the ground states is shown in Fig.  \ref{MottInsulator_Hubb}.   In our Mott insulator  particles are allowed to be localized both on lattice sites and on bonds as shown in Fig.  \ref{top_Mott:GS}.  

\begin{figure}[H]
\centering
\includegraphics[scale=0.12]{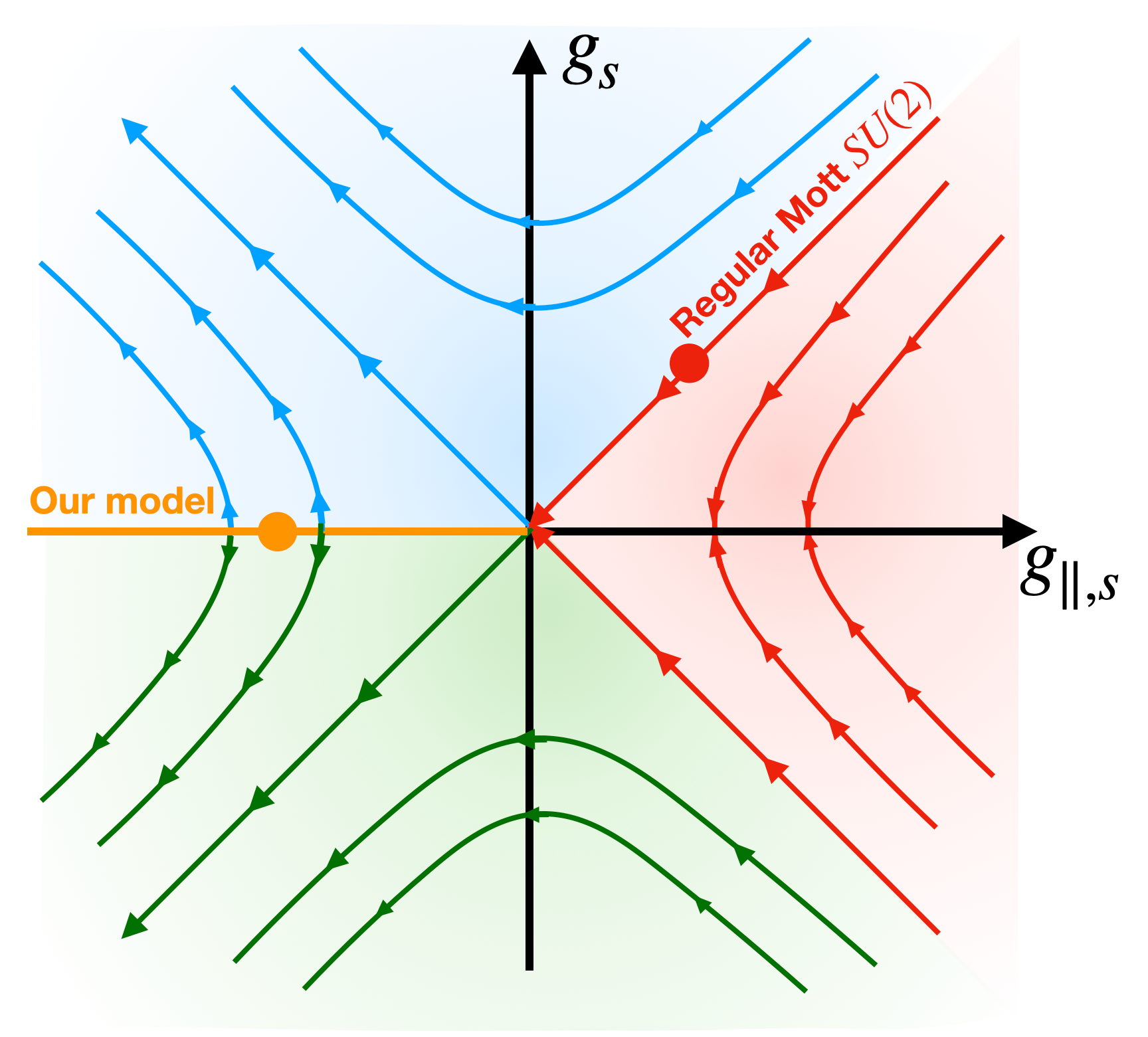}
\caption{RG flow of the model \eqref{full_model_main} in spin sector.  The trivial Mott insulator that occurs in our model is shown as an orange line.  Mott insulator of the Hubbard model occurs on the separatrix,  due to SU(2) symmetry of the model.  The two phases can be adiabatically connected,  by tuning the forward scattering amplitude $g_{\parallel,s}$ which is related to the Luttinger parameter in the spin sector}
\label{BKT}
\end{figure}

This difference arises due to the fact that our Mott phase occurs at the phase boundary between two gapped phases SDW and $\text{SSH}_+$,  where charges localized on sites or bonds,  while in the Hubbard model the Mott insulator is the phase that is protected by SU(2) symmetry and appears as a separatrix between gapless and gapped phases,  see Fig.  \ref{BKT}.  Even though the two ground states seem different,  physically they represent the same phase and thus can be adiabatically connected.  As shown in Fig. \ref{BKT},  tuning the forward-scattering amplitude allows one to move from the regular Mott phase on the SU(2)-symmetric line to our phase without reopening the gap.

\section{Gapless excitations in the generalized interacting model}\label{gapless}
Here we derive the excitation spectrum at the critical line $g_s=\delta t^2/16g_c$ in a general model with the gap-opening term \eqref{gap_full_deltat}.  At this critical line the fields in charge and spin sector are not fixed separately in the ground state of a model,  but are related by the following equation:
\begin{align}
\label{curves_minima}
\sqrt{2\pi}\phi_c = \pi  \pm  \arccos \left((\delta t/4g_c) \cos[ \sqrt{2\pi}\phi_s]\right) \mod 2\pi, 
\end{align} 
that describes two inequivalent curves of constant energy in the space of the two bosonic fields $(\phi_c, \phi_s)$.  Along those curves the spectrum is gapless.  In order to understand how the gapless excitations are related to the original charge and spin degrees of freedom,  we first need to compute the normalized tangential and normal vectors to the curves of constant energy \eqref{curves_minima}.    Let us denote the two vectors $\bf{t}$ and $\bf{n}$ correspondingly,  and focus on one of the two curves in (\ref{curves_minima}),  for example,  on $\sqrt{2\pi}\phi_c = \pi +\arccos \left((\delta t/4g_c) \cos[ \sqrt{2\pi}\phi_s]\right)$. 

 An arbitrary variation of the bosonic fields around these minima can be parametrized by a vector $\boldsymbol{\delta \phi}= (\delta \phi_c,\delta \phi_s)$.  The gapless excitations correspond to the case when $ \boldsymbol{\delta \phi} \parallel \bf{t} $ and the excitations in the normal direction $ \boldsymbol{\delta \phi} \parallel  \bf{n} $ are massive.   Let us write down the expression for the vectors $\bf{t}$ and $\bf{n}$ explicitly:

\begin{align}
\label{tn}
\textbf{\text{t}}= \left( \frac{\alpha \sin[\sqrt{2\pi} \phi_s]}{\sqrt{1+\alpha^2-2\alpha^2\cos^2[\sqrt{2\pi}\phi_s]}},\sqrt{\frac{1-\alpha^2\cos^2[\sqrt{2\pi}\phi_s]}{1+\alpha^2-2\alpha^2\cos^2[\sqrt{2\pi}\phi_s]}} \right) \equiv (t_c, t_s) , \\
\textbf{\text{n}}=\left(\sqrt{\frac{1-\alpha^2\cos^2[\sqrt{2\pi}\phi_s]}{1+\alpha^2-2\alpha^2\cos^2[\sqrt{2\pi}\phi_s]}},- \frac{\alpha \sin[\sqrt{2\pi} \phi_s]}{\sqrt{1+\alpha^2-2\alpha^2\cos^2[\sqrt{2\pi}\phi_s]}} \right) \equiv (n_c, n_s),
\end{align}
where $\alpha = (\delta t/4g_c)$.    
Using that we obtain that the massless excitations are characterised by the following Hamiltonian: 
\begin{align}
\label{H_parall}
H_{\parallel} = \left[ \frac{v_c}{2K_c} t_c^2 + \frac{v_s}{2K_s} t_s^2 \right] \cdot (\partial_x \phi_{\parallel})^2 + \left[ \frac{v_c K_c}{2} t_c^2 + \frac{v_s K_s}{2} t_s^2 \right] \cdot (\partial_x \theta_{\parallel})^2.
\end{align} 
Expressed in terms of charge and spin degrees of freedom,  these excitations are: 
\begin{align}
\label{phi_parall}
\delta \phi_c^{\parallel} = \delta \phi_{\parallel} \cdot t_c,  \hspace{0.5cm}
\delta \phi^{\parallel}_s =  \delta \phi_{\parallel} \cdot t_s.
\end{align}
For the gapped degrees of freedom we get the following Hamiltonian: 
\begin{align}
\label{H_perp}
H_{\perp} = \left[ \frac{v_c}{2K_c} n_c^2 + \frac{v_s}{2K_s} n_s^2 \right] \cdot (\partial_x \phi_{\perp})^2 + \left[ \frac{v_c K_c}{2} n_c^2 + \frac{v_s K_s}{2} n_s^2 \right] \cdot (\partial_x \theta_{\perp})^2 + \frac{m^2}{2} \phi^2_{\perp},
\end{align} 
where the mass is given by $m^2 = 4g_c (1-\alpha^2\cos[2\phi_s])$,  and the variations of charge and spin fields are: 
\begin{align}
\label{phi_perp}
\delta \phi^{\perp}_c =  \delta \phi_{\perp} \cdot  n_c,  \hspace{0.5cm}
\delta \phi^{\perp}_s =  \delta \phi_{\perp} \cdot n_s.
\end{align}
Let us now look at two interesting limits of these expressions, in particular in the limit $\alpha=0$ or $g_s = \delta t= 0$ we should reproduce the topological gapless phase where the spin sector is gapless and charge is gapped.  Indeed,  from (\ref{phi_parall}) and (\ref{phi_perp}) we see that in this limit $\delta \phi_{\parallel} = \delta \phi_s^{\parallel}$ and $\delta \phi_{\perp} = \delta\phi_c^{\perp}$.  We may obtain a similar result in the limit $g_c =0$ or $\alpha \rightarrow \infty$,  where we reproduce the gapless phase in the charge sector. Note that this limit needs a bit more careful treatment as one needs take into account that $\phi_s$ is fixed,  as follows from (\ref{curves_minima}).
Another interesting special case is $\alpha=1$ that corresponds to the point $g_c=g_s=\delta t/4$  in the phase diagram in Fig.  \ref{phase_deltat}.  We discuss this case in the main text and show how the model can be deformed to a non-interacting metal at this point.  For general values of $\alpha$ the excitations carry both charge and spin,  moreover note that both Hamiltonians for gapless excitations 
\eqref{H_parall} and for gapped ones \eqref{H_perp} contain higher-order terms that are non-linear in bosonic fields,  both in kinetic energy part and in the mass $m$.  Such terms,  however,  are less relevant in the RG sense than the kinetic part and the mass term,  and therefore can be neglected.

\bibliography{Manuscript_v1}
\end{document}